\documentclass[preprint, superscriptaddress, preprintnumbers, amsmath, amssymb]{revtex4}

\allowdisplaybreaks
\allowdisplaybreaks[4]

\usepackage{graphicx}
\usepackage{dcolumn}
\usepackage{bm}
\usepackage{multirow}
\usepackage[dvipdfm,
            pdfstartview=FitH,
            bookmarks=true,
            bookmarksnumbered=true,
            bookmarksopen=true,
            colorlinks=true,
            pdfborder=001,
            linkcolor=blue,
            anchorcolor=blue,
            citecolor=blue,
            ]{hyperref}

\begin{document}

\title{Effective field theory for collective rotations and vibrations of
triaxially deformed nuclei}

\author{Q. B. Chen}
\affiliation{Physik-Department, Technische Universit\"{a}t
M\"{u}nchen, D-85747 Garching, Germany}

\author{N. Kaiser}
\affiliation{Physik-Department, Technische Universit\"{a}t
M\"{u}nchen, D-85747 Garching, Germany}

\author{Ulf-G. Mei{\ss}ner}
\affiliation{Helmholtz-Institut f\"{u}r Strahlen- und Kernphysik
and Bethe Center for Theoretical Physics,
Universit\"{a}t Bonn, D-53115 Bonn, Germany}

\affiliation{Institute for Advanced Simulation, Institut f\"{u}r Kernphysik,
J\"{u}lich Center for Hadron Physics and JARA-HPC, Forschungszentrum J\"{u}lich,
D-52425 J\"{u}lich, Germany}

\author{J. Meng}
\affiliation{State Key Laboratory of Nuclear Physics and Technology,
             School of Physics, Peking University, Beijing 100871, China}%
\affiliation{Yukawa Institute for Theoretical Physics, Kyoto
             University, Kyoto 606-8502, Japan}
\affiliation{Department of Physics, University of Stellenbosch,
             Stellenbosch, South Africa}%

\date{\today}

\begin{abstract}

The effective field theory (EFT) for triaxially deformed even-even
nuclei is generalized to include the vibrational degrees of freedom.
The pertinent Hamiltonian is constructed up to next-to-leading order.
The leading order  part describes the
vibrational motion, and the NLO  part couples rotations to
vibrations. The applicability of the EFT Hamiltonian is examined
through the description of the energy spectra of the ground state bands,
$\gamma$-bands, and $K=4$ bands in the $^{108, 110, 112}$Ru isotopes.
It is found that by taking into account the
vibrational degrees of freedom, the deviations for high-spin states
in the $\gamma$-band observed in the EFT with only rotational degrees of
freedom disappear. This supports the importance of including
vibrational degrees of freedom in the EFT formulation for the
collective motion of triaxially deformed nuclei.

\end{abstract}


\maketitle



\section{Introduction}\label{sec1}

Effective field theory (EFT) is based
on symmetry principles alone, and it exploits the separation of
energy scales for the systematic construction of the Hamiltonian
supplemented by a power counting. In this way, an increase in the
number of parameters (i.e., low-energy constants that need to be
adjusted to data) goes hand in hand with an increase in precision
and thereby counter balances the partial loss of predictive power.
Actually, EFT often exhibits an impressive efficiency as highlighted
by analytical results and economical means of calculations. In
recent decades, EFT has enjoyed considerable successes in low-energy
hadronic and nuclear structure physics. Pertinent examples include
the descriptions of the nuclear interactions~\cite{Epelbaum2006PPNP,
Epelbaum2009RMP, Machleidt2011PR}, halo nuclei~\cite{Bertulani2002NPA,
Hammer2011NPA, Ryberg2014PRC}, and nuclear few-body
systems~\cite{Bedaque2002ARNPS, Griesshammer2012PPNP, Hammer2013RMP}.
Recently, Papenbrock and his collaborators have
developed an EFT to describe the collective rotational
and vibrational motions of deformed nuclei by~\cite{Papenbrock2011NPA,
J.L.Zhang2013PRC, Papenbrock2014PRC, Papenbrock2015JPG,
Papenbrock2016PS, Perez2015PRC, Perez2015PRC_v1, Perez2016PRC}.

Collective rotations and vibrations are the typical low-lying
excitation modes of a nucleus. For a spherical nucleus,
only vibrational modes exist. On the other hand, for a deformed nucleus,
various vibrational bands are observed, and rotational bands are found
to be built on the successive vibrational modes~\cite{Bohr1975, Ring1980book}.
Since the initial paper in 2011~\cite{Papenbrock2011NPA}, Papenbrock and his
collaborators have completed a series of works devoted to the systematic
treatment of nuclear collective motion in the EFT framework~\cite{J.L.Zhang2013PRC,
Papenbrock2014PRC, Papenbrock2015JPG, Papenbrock2016PS,
Perez2015PRC, Perez2015PRC_v1, Perez2016PRC}. Through
the application of EFT to deformed nuclei, the finer details of the
experimental energy spectra, such as the change of the moment of inertia
with spin, can be addressed properly through higher-order
correction terms~\cite{Papenbrock2011NPA, J.L.Zhang2013PRC,
Papenbrock2014PRC, Papenbrock2015JPG, Papenbrock2016PS}. Moreover, in
this approach the uncertainties of the theoretical model can be
quantified~\cite{Perez2015PRC_v1}, and a consistent treatment of
electroweak currents together with the Hamiltonian can be
obtained~\cite{Perez2015PRC}. Let us note that all these
investigations were restricted to axially symmetric nuclei.

Very recently, the pertinent EFT has been further generalized to describe
the rotational motion of triaxially deformed even-even nuclei~\cite{Q.B.Chen2017EPJA}.
The triaxial deformation of nucleus has been a subject of much interest in the theoretical
study of nuclear structure for a long time. It is related to many
interesting phenomena, including the $\gamma$-band~\cite{Bohr1975}, signature
inversion~\cite{Bengtsson1984NPA}, anomalous signature
splitting~\cite{Hamamoto1988PLB}, the wobbling
motion~\cite{Bohr1975}, chiral rotational
modes~\cite{Frauendorf1997NPA} and most prominently multiple
chiral doublet bands~\cite{J.Meng2006PRC}. In fact, the wobbling
motion and chiral doublet bands are regarded as unique
fingerprints of stable triaxially deformed nuclei. In Ref.~\cite{Q.B.Chen2017EPJA},
the pertinent Hamiltonian has been constructed up to next-to-leading order (NLO).
Taking the energy spectra of the ground state and $\gamma$- bands (together
with some $K=4$ bands) in the $^{102-112}$Ru isotopes as examples, the applicability
of this novel EFT for triaxial nuclei has been examined. It has been
found that the description at NLO is overall better than at leading order (LO).
Nevertheless, there were still some deviations between the NLO calculation
and the data for some high-spin states in the $\gamma$-bands. The comparison to
the results of a five-dimensional collective Hamiltonian (5DCH) based on the
covariant density functional theory (CDFT)~\cite{J.Meng2016book} has indicated
that the inclusion of vibrational degrees of freedom
in the EFT formulation is important.

Therefore, in this paper, the vibrational degrees of freedom are additionally
considered in the formulation of an EFT for collective nuclear motion.
The pertinent Hamiltonian will be constructed
up to NLO, where the LO part describes the vibrational motion,
and the NLO part couples rotations to vibrations. The energy spectra
of ground state bands, $\gamma$-bands, and $K=4$ bands in the $^{108,110,112}$Ru
isotopes are taken as examples to examine the applicability of our extended
EFT approach.

The present paper is organized as follows. In Sec.~\ref{sec2}, the EFT for collective
rotations and vibrations of triaxially deformed nuclei is constructed. The
solutions of the rotational Hamiltonian in first order perturbation theory
are given in Sec.~\ref{sec3}. The obtained vibrational
Hamiltonian is reduced in Sec.~\ref{sec4} by expressing it in
terms of the quadrupole deformation parameters $\beta_2$ and $\gamma_2$. The results
of the corresponding quantum-mechanical calculations are presented
and discussed in detail in Sec.~\ref{sec5}. Finally, a summary is given
in Sec.~\ref{sec6} together with perspectives for future research directions.


\section{Construction of the EFT}\label{sec2}

In this section, the procedure of constructing the effective Lagrangian
and Hamiltonian for collective rotations and vibrations
is introduced and carried out. It follows similar steps as in the case
of axially symmetric nuclei studied in Refs.~\cite{Papenbrock2014PRC, Papenbrock2015JPG}
and in the case of collective rotations of triaxially deformed nuclei
investigated in Ref.~\cite{Q.B.Chen2017EPJA}.

\subsection{Dynamical variables}

In an EFT, the symmetry is (typically) realized nonlinearly~\cite{Coleman1969PR,
Callan1969PR}, and the Nambu-Goldstone fields parametrize the coset space
$\mathcal{G}/\mathcal{H}$. Here, $\mathcal{G}$ is the symmetry group of the
Hamiltonian and $\mathcal{H}$ the symmetry group of the ground state,
is a proper subgroup of $\mathcal{G}$. The effective Lagrangian $\mathcal{L}$
is built from invariants that can be constructed from the fields in the coset space.
In the following, we will write the fields relevant for collective nuclear motion
in the space-fixed coordinate frame, where the three generators
of infinitesimal rotations about the space-fixed $x$, $y$,
and $z$-axes are denoted by $J_x$, $J_y$, and $J_z$.

To describe a global rotation, the three Euler angles $\alpha$, $\beta$, and $\gamma$
serve as natural dynamical variables. On the classical level, they are purely
time-dependent and upon quantization, they give rise to rotational bands.
On the other hand, vibrations act locally on the nuclear surface and its location can be
described by body-fixed spherical coordinates $r$, $\theta$, and $\phi$.
Following the arguments in Refs.~\cite{Papenbrock2014PRC, Papenbrock2015JPG, Papenbrock2016PS},
one expects that Nambu-Goldstone modes related to the radial coordinate $r$
have higher frequencies than those related to the angles $\theta$ and $\phi$.
For low energy excitations, one can therefore restrict the
attention to the angular variables.

A triaxially deformed nucleus is invariant under the $180^\circ$ rotation
about the body-fixed axes (discrete $\textrm{D}_2$ symmetry),
while the continuous $\textrm{SO}(3)$ symmetry is
broken by the deformation. Consequently, the Nambu-Goldstone modes
lie in the three-dimensional coset space $\textrm{SO}(3)/\textrm{D}_2$. The modes
depend on the angles $\theta$ and $\phi$ and the time $t$ and are
generated from the nuclear ground state by a unitary transformation $U$. Following
Refs.~\cite{Papenbrock2014PRC, Papenbrock2015JPG, Papenbrock2016PS, Q.B.Chen2017EPJA},
the matrix $U$ can be parameterized in a product form as
\begin{align}\label{chp5:eq6}
 U&=g(\alpha,\beta,\gamma)u(x,y,z),\notag\\
 g(\alpha,\beta,\gamma)&=\exp\{-i\alpha(t)J_z\}
     \exp\{-i\beta(t)J_y\}\exp\{-i\gamma(t)J_z\},\notag\\
 u(x,y,z)&=\exp\{-ix(\theta,\phi,t)J_x
     -iy(\theta,\phi,t)J_y-iz(\theta,\phi,t)J_z\}.
\end{align}
The fields $x(\theta, \phi, t)$, $y(\theta, \phi, t)$, and
$z(\theta, \phi, t)$ with $|x|$, $|y|$, $|z| \ll 1$ generate
small-amplitude vibrations of the nuclear surface. They depend
on $\theta$ and $\phi$ in such that their angular averages
vanish
\begin{align}
 \int d\Omega~x(\theta,\phi,t)=\int d\Omega~y(\theta,\phi,t)
  =\int d\Omega~z(\theta,\phi,t)=0,
\end{align}
where $d\Omega=\sin\theta\, d\theta d \phi$ denotes the surface
element on the unit sphere. Note that in the axially symmetric case~\cite{Papenbrock2014PRC,
Papenbrock2015JPG}, $\gamma(t)$ and $z(\theta,\phi,t)$ do not exist as
observable degrees of freedom, since the operator $J_z$ acting on the axially
deformed ground state gives zero.

The power counting underlying the EFT is specified by~\cite{Papenbrock2016PS}
\begin{align}
 & \alpha, \beta, \gamma \sim \mathcal{O}(1), \quad
 \dot{\alpha},\dot{\beta},\dot{\gamma}\sim \xi, \\
 & x, y, z \sim \sqrt{\xi/\Omega}, \quad
 \dot{x},\dot{y},\dot{z} \sim \sqrt{\xi\Omega},\quad
 \partial_\nu x, \partial_\nu y, \partial_\nu z \sim \sqrt{\xi/\Omega},
\end{align}
where the $\xi$ and the $\Omega$ denote the energy scales of the rotational and
vibrational motion, respectively. The dot refers to a time derivative and
$\partial_\nu$ to angular derivatives with $\nu=\theta$ or $\phi$. Note that $\xi$
(typically $\approx 80$ keV) is much smaller than $\Omega$ (typically $\approx$ 1
MeV), and hence $\sqrt{\xi/\Omega}\ll 1$ is a reasonably small parameter.

\subsection{Effective Lagrangian}

As usual, the effective Lagrangian is built from invariants. These are
constructed from quantities $a_\mu^x$, $a_\mu^y$, and $a_\mu^z$
defined by~\cite{Coleman1969PR, Callan1969PR, Papenbrock2014PRC,
Papenbrock2015JPG, Q.B.Chen2017EPJA}
\begin{align}\label{chp5:eq41}
 iU^{-1}\partial_\mu U=a_\mu^x J_x + a_\mu^y J_y + a_\mu^z J_z.
\end{align}
The symbol $\partial_\mu$ stands for a derivative with
respect to one of the variables $\theta$, $\phi$, and $t$.

To work out the Nambu-Goldstone modes explicitly, we use the
decompositions
\begin{align}
 ig^{-1}\partial_t g&=(-\dot{\alpha}\sin\beta\cos\gamma+\dot{\beta}\sin\gamma)J_x
 + (\dot{\alpha}\sin\beta\sin\gamma+\dot{\beta}\cos\gamma)J_y
 + (\dot{\alpha}\cos\beta+\dot{\gamma})J_z,
\end{align}
and
\begin{align}
 iu^{-1}\partial_t u
  &=\Big\{\dot{x}+\frac{1}{2}(\dot{y}z-y\dot{z})
   -\frac{1}{6}\Big[(y^2+z^2)\dot{x}-x(y\dot{y}+z\dot{z})\Big]\Big\}J_x\notag\\
  &\quad +\Big\{\dot{y}+\frac{1}{2}(\dot{z}x-z\dot{x})
   -\frac{1}{6}\Big[(x^2+z^2)\dot{y}-y(x\dot{x}+z\dot{z})\Big]\Big\}J_y\notag\\
  &\quad +\Big\{\dot{z}+\frac{1}{2}(\dot{x}y-x\dot{y})
   -\frac{1}{6}\Big[(x^2+y^2)\dot{z}-z(x\dot{x}+y\dot{y})\Big]\Big\}J_z~.
\end{align}
In the calculation of $i u^{-1}\partial_t u$, the expansion of $u$
in powers of $x$, $y$, $z$ has been performed up to cubic terms.
With these formula, one obtains for the time derivative
\begin{align}\label{chp5:eq17}
 iU^{-1}\partial_t U&=iu^{-1}\partial_t u+i u^{-1}(g^{-1}\partial_t g)u\notag\\
 &=\Big\{\dot{x}+\frac{1}{2}(\dot{y}z-y\dot{z})
   -\frac{1}{6}\Big[(y^2+z^2)\dot{x}-x(y\dot{y}+z\dot{z})\Big]
   +(-\dot{\alpha}\sin\beta\cos\gamma+\dot{\beta}\sin\gamma)\notag\\
 &\quad +(\dot{\alpha}\sin\beta\sin\gamma+\dot{\beta}\cos\gamma)z
   -(\dot{\alpha}\cos\beta+\dot{\gamma})y\Big\}J_x\notag\\
 &\quad +\Big\{\dot{y}+\frac{1}{2}(\dot{z}x-z\dot{x})
   -\frac{1}{6}\Big[(x^2+z^2)\dot{y}-y(x\dot{x}+z\dot{z})\Big]
   +(\dot{\alpha}\sin\beta\sin\gamma+\dot{\beta}\cos\gamma)\notag\\
 &\quad +(\dot{\alpha}\cos\beta+\dot{\gamma})x
   -(-\dot{\alpha}\sin\beta\cos\gamma+\dot{\beta}\sin\gamma)z\Big\}J_y\notag\\
 &\quad +\Big\{\dot{z}+\frac{1}{2}(\dot{x}y-x\dot{y})
   -\frac{1}{6}\Big[(x^2+y^2)\dot{z}-z(x\dot{x}+y\dot{y})\Big]
   +(\dot{\alpha}\cos\beta+\dot{\gamma})\notag\\
 &\quad +(-\dot{\alpha}\sin\beta\cos\gamma+\dot{\beta}\sin\gamma)y
   -(\dot{\alpha}\sin\beta\sin\gamma+\dot{\beta}\cos\gamma)x\Big\}J_z,
\end{align}
while the angular derivatives ($\nu=\theta,\phi$) read
\begin{align}\label{chp5:eq18}
 iU^{-1}\partial_\nu U&=iu^{-1}\partial_\nu u\notag\\
  &=\Big\{\partial_\nu x+\frac{1}{2}(z\partial_\nu y-y\partial_\nu z)
   -\frac{1}{6}\Big[(y^2+z^2)\partial_\nu x-x(y\partial_\nu y+z\partial_\nu z)\Big]\Big\}J_x\notag\\
  &\quad +\Big\{\partial_\nu y+\frac{1}{2}(x\partial_\nu z-z\partial_\nu x)
   -\frac{1}{6}\Big[(x^2+z^2)\partial_\nu y-y(x\partial_\nu x+z\partial_\nu z)\Big]\Big\}J_y\notag\\
  &\quad +\Big\{\partial_\nu z+\frac{1}{2}(y\partial_\nu x-x\partial_\nu y)
   -\frac{1}{6}\Big[(x^2+y^2)\partial_\nu z-z(x\partial_\nu x+y\partial_\nu y)\Big]\Big\}J_z.
\end{align}
In the expansions of $iU^{-1}\partial_t U$ in Eq.~(\ref{chp5:eq17})
we go up to order $\xi\sqrt{\xi/\Omega}$ to include the coupling terms between
rotations and vibrations. Therefore, the quantities $u^{-1}J_xu$,
$u^{-1}J_yu$, and $u^{-1}J_zu$ need to be expanded only up to order
$\sqrt{\xi/\Omega}$, since $g^{-1}\partial_t g$ is of order $\xi$.

The components of the angular velocities follow from Eqs.~(\ref{chp5:eq41})
and (\ref{chp5:eq17}) as
\begin{align}
\label{chp5:eq42}
 a_t^x&=\dot{x}+\Big[\frac{1}{2}(\dot{y}z-y\dot{z})
   +(-\dot{\alpha}\sin\beta\cos\gamma+\dot{\beta}\sin\gamma)\Big]
   +\Big\{-\frac{1}{6}\Big[(y^2+z^2)\dot{x}-x(y\dot{y}+z\dot{z})\Big]\notag\\
 &\quad +(\dot{\alpha}\sin\beta\sin\gamma+\dot{\beta}\cos\gamma)z
   -(\dot{\alpha}\cos\beta+\dot{\gamma})y\Big\},\\
 a_t^y&=\dot{y}+\Big[\frac{1}{2}(\dot{z}x-z\dot{x})
   +(\dot{\alpha}\sin\beta\sin\gamma+\dot{\beta}\cos\gamma)\Big]
   +\Big\{-\frac{1}{6}\Big[(x^2+z^2)\dot{y}-y(x\dot{x}+z\dot{z})\Big]\notag\\
  &\quad +(\dot{\alpha}\cos\beta+\dot{\gamma})x
   -(-\dot{\alpha}\sin\beta\cos\gamma+\dot{\beta}\sin\gamma)z\Big\},\\
 a_t^z&=\dot{z}+\Big[\frac{1}{2}(\dot{x}y-x\dot{y})
   +(\dot{\alpha}\cos\beta+\dot{\gamma})\Big]
   +\Big\{-\frac{1}{6}\Big[(x^2+y^2)\dot{z}-z(x\dot{x}+y\dot{y})\Big]\notag\\
  &\quad +(-\dot{\alpha}\sin\beta\cos\gamma+\dot{\beta}\sin\gamma)y
   -(\dot{\alpha}\sin\beta\sin\gamma+\dot{\beta}\cos\gamma)x\Big\}~,
\end{align}
where the leading terms ($\dot{x}$, $\dot{y}$, and $\dot{z}$) are of order
$\sqrt{\xi\Omega}$. The terms in square brackets are proportional to the
energy scale $\xi$, and those in curly brackets scale as $\xi\sqrt{\xi/\Omega}$.
Thus, successive terms in $a_t^{x,y,z}$ are suppressed by a relative
factor $\sqrt{\xi/\Omega}$.

Moreover, one obtains from Eqs.~(\ref{chp5:eq41}) and (\ref{chp5:eq18}) for the
components involving angular derivatives
\begin{align}
\label{chp5:eq10}
 a_\nu^x&=\partial_\nu x+\frac{1}{2}(z\partial_\nu y-y\partial_\nu z)
   -\frac{1}{6}\Big[(y^2+z^2)\partial_\nu x-x(y\partial_\nu y+z\partial_\nu z)\Big],\\
\label{chp5:eq11}
 a_\nu^y&=\partial_\nu y+\frac{1}{2}(x\partial_\nu z-z\partial_\nu x)
   -\frac{1}{6}\Big[(x^2+z^2)\partial_\nu y-y(x\partial_\nu x+z\partial_\nu z)\Big],\\
\label{chp5:eq12}
 a_\nu^z&=\partial_\nu z+\frac{1}{2}(y\partial_\nu x-x\partial_\nu y)
   -\frac{1}{6}\Big[(x^2+y^2)\partial_\nu z-z(x\partial_\nu x+y\partial_\nu y)\Big],
\end{align}
where the leading terms ($\partial_\nu x$, $\partial_\nu y$, and $\partial_\nu z$)
scale as $\sqrt{\xi/\Omega}$, the terms in round brackets as $\xi/\Omega$,
and those square brackets as $(\xi/\Omega)^{3/2}$. Again, successive terms
in $a_\nu^{x,y,z}$ are suppressed by a relative factor $\sqrt{\xi/\Omega}$.

The expressions in Eqs.~(\ref{chp5:eq42})-(\ref{chp5:eq12}) give the lowest
order contributions from the Nambu-Goldstone modes $x, y, z$ and $\alpha, \beta, \gamma$
introduced in Eq.~(\ref{chp5:eq6}). Note that if $\gamma=0$ and $z=0$, then
one recovers the axailly symmetric case~\cite{Papenbrock2014PRC, Papenbrock2015JPG}.

Respecting time-reversal invariance and the discrete $D_2$ symmetry of a triaxial
nucleus~\footnote{Under the four elements of $\textrm{D}_{2}=\textrm{Z}_2\times \textrm{Z}_2$
the angular velocity vector $(a_t^x, a_t^y, a_t^z)$ is transformed into $(a_t^x, a_t^y, a_t^z)$,
$(a_t^x, -a_t^y, -a_t^z)$, $(-a_t^x, a_t^y, -a_t^z)$, and
$(-a_t^x, -a_t^y, a_t^z)$, respectively.}, only the squares of
$a_\mu^{x}$, $a_\mu^{y}$, and $a_\mu^{z}$ may be used to construct the
Lagrangian. For the temporal parts, one gets
\begin{align}
\label{chp5:eq14}
 (a_t^x)^2
  &=(-\dot{\alpha}\sin\beta\cos\gamma+\dot{\beta}\sin\gamma)^2\notag\\
  &\quad +\dot{x}^2
   +(\dot{y}z-y\dot{z})(-\dot{\alpha}\sin\beta\cos\gamma+\dot{\beta}\sin\gamma)\notag\\
  &\quad +2(\dot{\alpha}\sin\beta\sin\gamma+\dot{\beta}\cos\gamma)\dot{x}z
   -2(\dot{\alpha}\cos\beta+\dot{\gamma})\dot{x}y\notag\\
  &\quad +\frac{1}{4}(\dot{y}z-y\dot{z})^2
   -\frac{1}{3}\Big[(y^2+z^2)\dot{x}^2-x\dot{x}(y\dot{y}+z\dot{z})\Big],\\
\label{chp5:eq15}
 (a_t^y)^2
  &=(\dot{\alpha}\sin\beta\sin\gamma+\dot{\beta}\cos\gamma)^2\notag\\
  &\quad +\dot{y}^2
   +(\dot{z}x-z\dot{x})(\dot{\alpha}\sin\beta\sin\gamma+\dot{\beta}\cos\gamma)\notag\\
  &\quad +2(\dot{\alpha}\cos\beta+\dot{\gamma})\dot{y}x
   -2(-\dot{\alpha}\sin\beta\cos\gamma+\dot{\beta}\sin\gamma)\dot{y}z\notag\\
  &\quad +\frac{1}{4}(\dot{z}x-z\dot{x})^2
   -\frac{1}{3}\Big[(x^2+z^2)\dot{y}^2-y\dot{y}(x\dot{x}+z\dot{z})\Big],\\
\label{chp5:eq16}
 (a_t^z)^2
  &=(\dot{\alpha}\cos\beta+\dot{\gamma})^2\notag\\
  &\quad +\dot{z}^2
   +(\dot{x}y-x\dot{y})(\dot{\alpha}\cos\beta+\dot{\gamma})\notag\\
  &\quad +2(-\dot{\alpha}\sin\beta\cos\gamma+\dot{\beta}\sin\gamma)\dot{z}y
   -2(\dot{\alpha}\sin\beta\sin\gamma+\dot{\beta}\cos\gamma)\dot{z}x\notag\\
  &\quad +\frac{1}{4}(\dot{x}y-x\dot{y})^2
   -\frac{1}{3}\Big[(x^2+y^2)\dot{z}^2-z\dot{z}(x\dot{x}+y\dot{y})\Big],
\end{align}
where terms of order $\xi\Omega(\xi/\Omega)^{3/2}$ and higher have been
consistently dropped. The invariants in Eqs.~(\ref{chp5:eq14})-(\ref{chp5:eq16})
can still be decomposed according to the power of the vibrational fields
into
\begin{align}
 \mathcal{L}_{1a}&=(-\dot{\alpha}\sin\beta\cos\gamma+\dot{\beta}\sin\gamma)^2,\\
 \mathcal{L}_{2a}&=\dot{x}^2+(\dot{y}z-y\dot{z})
      (-\dot{\alpha}\sin\beta\cos\gamma+\dot{\beta}\sin\gamma)\notag\\
   &\quad +2(\dot{\alpha}\sin\beta\sin\gamma+\dot{\beta}\cos\gamma)\dot{x}z
    -2(\dot{\alpha}\cos\beta+\dot{\gamma})\dot{x}y,\\
 \mathcal{L}_{3a}&=\frac{1}{4}(\dot{y}z-y\dot{z})^2
   -\frac{1}{3}\Big[(y^2+z^2)\dot{x}^2-x\dot{x}(y\dot{y}+z\dot{z})\Big],\\
 \mathcal{L}_{1b}&=(\dot{\alpha}\sin\beta\sin\gamma+\dot{\beta}\cos\gamma)^2,\\
 \mathcal{L}_{2b}&=\dot{y}^2+(\dot{z}x-z\dot{x})
      (\dot{\alpha}\sin\beta\sin\gamma+\dot{\beta}\cos\gamma)\notag\\
   &\quad +2(\dot{\alpha}\cos\beta+\dot{\gamma})\dot{y}x
    -2(-\dot{\alpha}\sin\beta\cos\gamma+\dot{\beta}\sin\gamma)\dot{y}z,\\
 \mathcal{L}_{3b}&=\frac{1}{4}(\dot{z}x-z\dot{x})^2
   -\frac{1}{3}\Big[(x^2+z^2)\dot{y}^2-y\dot{y}(x\dot{x}+z\dot{z})\Big],\\
 \mathcal{L}_{1c}&=(\dot{\alpha}\cos\beta+\dot{\gamma})^2,\\
 \mathcal{L}_{2c}&=\dot{z}^2+(\dot{x}y-x\dot{y})
      (\dot{\alpha}\cos\beta+\dot{\gamma})\notag\\
   &\quad +2(-\dot{\alpha}\sin\beta\cos\gamma+\dot{\beta}\sin\gamma)\dot{z}y
    -2(\dot{\alpha}\sin\beta\sin\gamma+\dot{\beta}\cos\gamma)\dot{z}x,\\
 \mathcal{L}_{3c}&=\frac{1}{4}(\dot{x}y-x\dot{y})^2
   -\frac{1}{3}\Big[(x^2+y^2)\dot{z}^2-z\dot{z}(x\dot{x}+y\dot{y})\Big],
\end{align}
where $\mathcal{L}_{1a, 3a}$, $\mathcal{L}_{1b, 3b}$, and
$\mathcal{L}_{1c, 3c}$ are of order $\xi^2$. The leading terms
($\dot{x}^2$, $\dot{y}^2$, and $\dot{z}^2$) in $\mathcal{L}_{2a, 2b, 2c}$ are
of order $\xi\Omega$ and the remaining terms
scale as order $\xi^2$.

Next, we turn to the invariants constructed from derivatives with
respect to the angles $\theta$ and $\phi$ listed in Eqs.~(\ref{chp5:eq10})-(\ref{chp5:eq12}).
We restrict ourselves to terms of up to fourth order in $x$, $y$, $z$,
and their derivatives. The pertinent squares read
\begin{align}
\label{chp5:eq38}
 (a_\nu^x)^2
  &=(\partial_\nu x)^2
   +\Big[(\partial_\nu x)(z\partial_\nu y-y\partial_\nu z)\Big]\notag\\
  &\quad +\Big\{\frac{1}{4}(z\partial_\nu y-y\partial_\nu z)^2
   -\frac{1}{3}(\partial_\nu x)
   \Big[(y^2+z^2)\partial_\nu x-x(y\partial_\nu y+z\partial_\nu z)\Big]\Big\},\\
\label{chp5:eq39}
 (a_\nu^y)^2
  &=(\partial_\nu y)^2
   +\Big[(\partial_\nu y)(x\partial_\nu z-z\partial_\nu x)\Big]\notag\\
  &\quad +\Big\{\frac{1}{4}(x\partial_\nu z-z\partial_\nu x)^2
   -\frac{1}{3}(\partial_\nu y)
   \Big[(x^2+z^2)\partial_\nu y-y(x\partial_\nu x+z\partial_\nu z)\Big]\Big\},\\
\label{chp5:eq40}
 (a_\nu^z)^2
  &=(\partial_\nu z)^2
   +\Big[(\partial_\nu z)(y\partial_\nu x-x\partial_\nu y)\Big]\notag\\
  &\quad +\Big\{\frac{1}{4}(y\partial_\nu x-x\partial_\nu y)^2
   -\frac{1}{3}(\partial_\nu z)
   \Big[(x^2+y^2)\partial_\nu z-z(x\partial_\nu x+y\partial_\nu y)\Big]\Big\},
\end{align}
where terms of order $(\xi/\Omega)^{5/2}$ and higher have been consistently
dropped. In the above, $(a_\nu^x)^2$, $(a_\nu^y)^2$, and $(a_\nu^z)^2$
are expressed in terms of $x$, $y$, $z$, and their derivatives as
polynomials of degree two, three, and four, respectively.

The partial derivatives $\partial_\nu$ can be replaced by
the orbital angular momenta operators $\bm{L}$, whose components are
\begin{align}
 L_x&=i(\sin\phi~\partial_\theta+\cot\theta\cos\phi~\partial_\phi),\\
 L_y&=i(-\cos\phi~\partial_\theta+\cot\theta\sin\phi~\partial_\phi),\\
 L_z&=-i\partial_\phi.
\end{align}
By reexpressing $\partial_\theta$ and $\partial_\phi$ in terms of $L_x$,
$L_y$, and $L_z$, one constructs from the first terms in
Eqs.~(\ref{chp5:eq38})-(\ref{chp5:eq40}) the following six Lagrangians
\begin{align}
 \mathcal{L}_{4a}&=(\bm{L}x)^2, \quad
 \mathcal{L}_{4b}=(\bm{L}y)^2, \quad
 \mathcal{L}_{4c}=(\bm{L}z)^2,\\
 \mathcal{L}_{4d}&=(L_z x)^2, \quad
 \mathcal{L}_{4e}=(L_z y)^2, \quad
 \mathcal{L}_{4f}=(L_z z)^2.
\end{align}
In the same way, the second terms in square brackets lead to
\begin{align}
 \mathcal{L}_{5a}&=(\bm{L}x)\big[z(\bm{L}y)-y(\bm{L}z)\big],\\
 \mathcal{L}_{5b}&=(\bm{L}y)\big[x(\bm{L}z)-z(\bm{L}x)\big],\\
 \mathcal{L}_{5c}&=(\bm{L}z)\big[y(\bm{L}x)-x(\bm{L}y)\big],\\
 \mathcal{L}_{5d}&=(L_y x)\big[z(L_x y)-y(L_x z)\big]+(L_x x)\big[z(L_y y)-y(L_y z)\big],\\
 \mathcal{L}_{5e}&=(L_y y)\big[x(L_x z)-z(L_x x)\big]+(L_x y)\big[x(L_y z)-z(L_y x)\big],\\
 \mathcal{L}_{5f}&=(L_y z)\big[y(L_x x)-x(L_x y)\big]+(L_x z)\big[y(L_y x)-x(L_y y)\big],
\end{align}
and the third terms in curly brackets give rise to
\begin{align}
 \mathcal{L}_{6a}&=\frac{1}{4}\big[z(\bm{L}y)-y(\bm{L}z)\big]^2
  +\frac{1}{3}(\bm{L} x)\big\{x\big[y(\bm{L} y)+z(\bm{L} z)\big]-(y^2+z^2)(\bm{L} x)\big\},\\
 \mathcal{L}_{6b}&=\frac{1}{4}\big[x(\bm{L}z)-z(\bm{L}x)\big]^2
  +\frac{1}{3}(\bm{L} y)\big\{y\big[z(\bm{L} z)+x(\bm{L} x)\big]-(z^2+x^2)(\bm{L} y)\big\},\\
 \mathcal{L}_{6c}&=\frac{1}{4}\big[y(\bm{L}x)-x(\bm{L}y)\big]^2
  +\frac{1}{3}(\bm{L} z)\big\{z\big[x(\bm{L} x)+y(\bm{L} y)\big]-(x^2+y^2)(\bm{L} z)\big\},\\
 \mathcal{L}_{6d}&=\frac{1}{2}\big[z(L_y y)-y(L_y z)\big]\big[z(L_x y)-y(L_x z)\big]
  +\frac{1}{3}x\big\{(L_y x)\big[y(L_x y)+z(L_x z)\big]\notag\\
 &\quad +(L_x x)\big[y(L_y y)+z(L_y z)\big] \big\}-\frac{2}{3}(y^2+z^2)(L_x x)(L_y x),\\
 \mathcal{L}_{6e}&=\frac{1}{2}\big[x(L_y z)-z(L_y x)\big]\big[x(L_x z)-z(L_x x)\big]
  +\frac{1}{3}y\big\{(L_y y)\big[z(L_x z)+x(L_x x)\big]\notag\\
 &\quad +(L_x y)\big[z(L_y z)+x(L_y x)\big] \big\}-\frac{2}{3}(z^2+x^2)(L_x y)(L_y y),\\
 \mathcal{L}_{6f}&=\frac{1}{2}\big[y(L_y x)-x(L_y y)\big]\big[y(L_x x)-x(L_x y)\big]
  +\frac{1}{3}z\big\{(L_y z)\big[x(L_x x)+y(L_x y)\big]\notag\\
 &\quad +(L_x z)\big[x(L_y x)+y(L_y y)\big] \big\}-\frac{2}{3}(x^2+y^2)(L_x z)(L_y z).
\end{align}

As a result, the total effective Lagrangian $L$ is given by the angular
average of a arbitrary linear combination of the invariants constructed above and it
involves 27 low-energy constants (LECs)
\begin{align}\label{chp5:eq43}
 L&=L_1+L_2+L_3+L_4+L_5+L_6\notag\\
  &=\frac{1}{4\pi}\int d\Omega \Big[\sum_{i=a,b,c}\Big(\frac{A_i}{2}\mathcal{L}_{1i}
   +\frac{B_i}{2}\mathcal{L}_{2i}+\frac{C_i}{2}\mathcal{L}_{3i}\Big)
   -\sum_{i=a,b,c,d,e,f}\Big(\frac{D_i}{2}\mathcal{L}_{4i}
   +\frac{E_i}{2}\mathcal{L}_{5i}+\frac{F_i}{2}\mathcal{L}_{6i}\Big)\Big].
\end{align}

Following Ref.~\cite{Papenbrock2014PRC}, we expand the real function
$x(\theta,\phi,t)$ in terms of the real orthonormal functions $Z_{\lambda\mu}$,
called tesseral harmonics, as
\begin{align}\label{chp5:eq31}
 x(\theta,\phi,t)=\sum_{\lambda=2}^\infty
 \sum_{\mu=-\lambda}^\lambda x_{\lambda\mu}(\theta,\phi,t)Z_{\lambda\mu}(\theta,\phi),
\end{align}
and in the same way for the real variables $y$, $z$, $\dot{x}$, $\dot{y}$,
and $\dot{z}$. The tesseral harmonics $Z_{\lambda\mu}(\theta,\phi)$ are
related to the spherical harmonics $Y_{\lambda\mu}(\theta,\phi)$ by
\begin{align}
 Z_{\lambda\mu}
  =\left\{
  \begin{array}{ll}
   \displaystyle \frac{1}{\sqrt{2}}\Big(Y_{\lambda\mu}+Y_{\lambda\mu}^*\Big), & \mu>0, \\
    Y_{\lambda0}, & \mu=0,\\
   \displaystyle -\frac{i}{\sqrt{2}}\Big(Y_{\lambda\mu}-Y_{\lambda\mu}^*\Big), & \mu<0. \\
  \end{array}
  \right.
\end{align}
It is obvious that the expansion coefficients $x_{\lambda\mu}$ are real.
Note that in Eq.~(\ref{chp5:eq31}) the contributions with $\lambda=0$ and $\lambda=1$
must be excluded, since $\lambda = 0$ describes global rotations while
$\lambda = 1$ describes global translations in space.

Using these expansions and carrying out the solid angle integration,
the total Lagrangian takes the form
\begin{align}\label{chp5:eq19}
 L&=\frac{A_a}{2}\omega_x^2
   +\frac{A_b}{2}\omega_y^2
   +\frac{A_c}{2}\omega_z^2\notag\\
  &\quad +\frac{B_a}{2}\sum_{\lambda\mu}\Big[\dot{x}_{\lambda\mu}^2
   +(\dot{y}_{\lambda\mu}z_{\lambda\mu}-y_{\lambda\mu}\dot{z}_{\lambda\mu})\omega_x
   +2\omega_y\dot{x}_{\lambda\mu}z_{\lambda\mu}
   -2\omega_z\dot{x}_{\lambda\mu}y_{\lambda\mu}\Big]\notag\\
  &\quad +\frac{B_b}{2}\sum_{\lambda\mu}\Big[\dot{y}_{\lambda\mu}^2
   +(\dot{z}_{\lambda\mu}x_{\lambda\mu}-z_{\lambda\mu}\dot{x}_{\lambda\mu})\omega_y
   +2\omega_z\dot{y}_{\lambda\mu}x_{\lambda\mu}
   -2\omega_x\dot{y}_{\lambda\mu}z_{\lambda\mu}\Big]\notag\\
  &\quad +\frac{B_c}{2}\sum_{\lambda\mu}\Big[\dot{z}_{\lambda\mu}^2
   +(\dot{x}_{\lambda\mu}y_{\lambda\mu}-x_{\lambda\mu}\dot{y}_{\lambda\mu})\omega_z
   +2\omega_x\dot{z}_{\lambda\mu}y_{\lambda\mu}
   -2\omega_y\dot{z}_{\lambda\mu}x_{\lambda\mu}\Big]\notag\\
  &\quad -\frac{1}{2}\sum_{\lambda\mu} \Big\{\lambda(\lambda+1)\Big[D_a x_{\lambda\mu}^2
   +D_b y_{\lambda\mu}^2+D_c z_{\lambda\mu}^2\Big]+\mu^2\Big[D_d x_{\lambda\mu}^2
   +D_e y_{\lambda\mu}^2+D_f z_{\lambda\mu}^2\Big]\Big\}.
\end{align}
where we have restricted ourselves to terms up to orders with
$\xi$, $\Omega$, and $\sqrt{\xi\Omega}$. In this way the number of
LECs get reduced to 12. The pure rotor part
(first line in Eq.~(\ref{chp5:eq19})) is written in terms of
squares of
\begin{align}
 \omega_x&=-\dot{\alpha}\sin\beta\cos\gamma+\dot{\beta}\sin\gamma,\\
 \omega_y&= \dot{\alpha}\sin\beta\sin\gamma+\dot{\beta}\cos\gamma,\\
 \omega_z&= \dot{\alpha}\cos\beta+\dot{\gamma},
\end{align}
which are the rotational frequencies (in the body-fixed frame)
expressed through Euler angles and their time derivatives~\cite{Q.B.Chen2017EPJA}.

Finally, we study the dependence of the parameters on
the energy scales. Since $\omega_{x,y,z}\sim\xi$, one requires
$A_{a,b,c} \sim \xi^{-1}$ so that the rotational energy scales as
order $\xi$. Similarly, $\dot{x}_{\lambda\mu}$, $\dot{y}_{\lambda\mu}$,
$\dot{z}_{\lambda\mu}\sim\sqrt{\xi\Omega}$ leads to $B_{a,b,c} \sim \xi^{-1}$
so that the vibrational energy scales as order $\Omega$. This
implies that the rotation-vibration coupling term is of order
$\xi^{-1}\sqrt{\xi\Omega}\sqrt{\xi/\Omega}\xi=\xi$.
In addition, the scaling behavior $D_{a-f} \sim \Omega^2/\xi$
implies that the collective potential (last line in Eq.~(\ref{chp5:eq19}))
is of order $\Omega$.

\subsection{Canonical momenta}

From the Lagrangian (\ref{chp5:eq19}), one derives the following
canonical momenta
\begin{align}
\label{chp5:eq20}
 p_\alpha
  &=\frac{\partial L}{\partial \dot{\alpha}}
   =-\sin\beta\cos\gamma\Big[A_a \omega_x+\frac{B_a}{2}\sum_{\lambda\mu}
   (\dot{y}_{\lambda\mu}z_{\lambda\mu}-y_{\lambda\mu}\dot{z}_{\lambda\mu})
   -B_b\sum_{\lambda\mu}\dot{y}_{\lambda\mu}z_{\lambda\mu}
   +B_c\sum_{\lambda\mu}\dot{z}_{\lambda\mu}y_{\lambda\mu}\Big]\notag\\
  &\quad +\sin\beta\sin\gamma\Big[A_b \omega_y+B_a\sum_{\lambda\mu}\dot{x}_{\lambda\mu}z_{\lambda\mu}
   +\frac{B_b}{2}\sum_{\lambda\mu}
   (\dot{z}_{\lambda\mu}x_{\lambda\mu}-z_{\lambda\mu}\dot{x}_{\lambda\mu})
   -B_c\sum_{\lambda\mu}\dot{z}_{\lambda\mu}x_{\lambda\mu}\Big]\notag\\
  &\quad +\cos\beta\Big[A_c \omega_z-B_a\sum_{\lambda\mu}\dot{x}_{\lambda\mu}y_{\lambda\mu}
   +B_b\sum_{\lambda\mu}\dot{y}_{\lambda\mu}x_{\lambda\mu}
   +\frac{B_c}{2}\sum_{\lambda\mu}
   (\dot{x}_{\lambda\mu}y_{\lambda\mu}-x_{\lambda\mu}\dot{y}_{\lambda\mu}) \Big],\\
\label{chp5:eq21}
 p_\beta
  &=\frac{\partial L}{\partial \dot{\beta}}
   =\sin\gamma\Big[A_a\omega_x+\frac{B_a}{2}\sum_{\lambda\mu}
   (\dot{y}_{\lambda\mu}z_{\lambda\mu}-y_{\lambda\mu}\dot{z}_{\lambda\mu})
   -B_b\sum_{\lambda\mu}\dot{y}_{\lambda\mu}z_{\lambda\mu}
   +B_c\sum_{\lambda\mu}\dot{z}_{\lambda\mu}y_{\lambda\mu} \Big]\notag\\
  &\quad +\cos\gamma\Big[A_b\omega_y+B_a\sum_{\lambda\mu}\dot{x}_{\lambda\mu}z_{\lambda\mu}
   +\frac{B_b}{2}\sum_{\lambda\mu}
   (\dot{z}_{\lambda\mu}x_{\lambda\mu}-z_{\lambda\mu}\dot{x}_{\lambda\mu})
   -B_c\sum_{\lambda\mu}\dot{z}_{\lambda\mu}x_{\lambda\mu}\Big],\\
\label{chp5:eq22}
 p_\gamma
  &=\frac{\partial L}{\partial \dot{\gamma}}
   =A_c \omega_z-B_a\sum_{\lambda\mu}\dot{x}_{\lambda\mu}y_{\lambda\mu}
   +B_b\sum_{\lambda\mu}\dot{y}_{\lambda\mu}x_{\lambda\mu}
   +\frac{B_c}{2}\sum_{\lambda\mu}(\dot{x}_{\lambda\mu}y_{\lambda\mu}
   -x_{\lambda\mu}\dot{y}_{\lambda\mu}),\\
\label{chp5:eq23}
 p_{\lambda\mu}^x
  &=\frac{\partial L}{\partial \dot{x}_{\lambda\mu}}
   =B_a\Big[\dot{x}_{\lambda\mu}
   +\omega_y z_{\lambda\mu}-\omega_z y_{\lambda\mu}\Big]
   -\frac{B_b}{2}z_{\lambda\mu}\omega_y
   +\frac{B_c}{2}y_{\lambda\mu}\omega_z,\\
\label{chp5:eq24}
 p_{\lambda\mu}^y
  &=\frac{\partial L}{\partial \dot{y}_{\lambda\mu}}
   =\frac{B_a}{2}z_{\lambda\mu}\omega_x
   +B_b\Big[\dot{y}_{\lambda\mu}
   +\omega_zx_{\lambda\mu}-\omega_xz_{\lambda\mu}\Big]
   -\frac{B_c}{2}x_{\lambda\mu}\omega_z,\\
\label{chp5:eq25}
 p_{\lambda\mu}^z
  &=\frac{\partial L}{\partial \dot{z}_{\lambda\mu}}
   =-\frac{B_a}{2}y_{\lambda\mu}\omega_x
   +\frac{B_b}{2}x_{\lambda\mu}\omega_y
   +B_c\Big[\dot{z}_{\lambda\mu}
   +\omega_x y_{\lambda\mu}-\omega_y x_{\lambda\mu}\Big].
\end{align}
Given the expressions of the canonical momenta $p_\alpha$, $p_\beta$,
and $p_\gamma$, one obtains (by forming appropriate linear combinations)
the components of angular momentum along the principal axes in the body-fixed
frame
\begin{align}\label{chp5:eq44}
 I_1&=A_a \omega_x+\frac{B_a-2B_b}{2}\sum_{\lambda\mu}\dot{y}_{\lambda\mu}z_{\lambda\mu}
   +\frac{2B_c-B_a}{2}\sum_{\lambda\mu}y_{\lambda\mu}\dot{z}_{\lambda\mu},\\
 I_2&=A_b \omega_y+\frac{B_b-2B_c}{2}\sum_{\lambda\mu}\dot{z}_{\lambda\mu}x_{\lambda\mu}
   +\frac{2B_a-B_b}{2}\sum_{\lambda\mu}z_{\lambda\mu}\dot{x}_{\lambda\mu},\\
 \label{chp5:eq45}
 I_3&=A_c \omega_z+\frac{B_c-2B_a}{2}\sum_{\lambda\mu}\dot{x}_{\lambda\mu}y_{\lambda\mu}
   +\frac{2B_b-B_c}{2}\sum_{\lambda\mu}x_{\lambda\mu}\dot{y}_{\lambda\mu}.
\end{align}
One observes that both rotation and vibration contribute to the angular
momentum. For the rotation, these are the usual products of moment of inertia and
rotational frequency $\omega_x$, $\omega_y$, or $\omega_z$.
For the vibration, the additional angular momentum arises from the motion of
the nuclear surface with (anisotropic) mass parameters $B_{a,b,c}$.

In the expressions for $p_{\lambda\mu}^x$, $p_{\lambda\mu}^y$,
and $p_{\lambda\mu}^z$, terms such as $\dot{x}_{\lambda\mu}$
are of order $\sqrt{\xi\Omega}$, while cross terms such as
$\omega_y z_{\lambda\mu}$ are of order $\xi\sqrt{\xi/\Omega}$.
Since the latter ones are suppressed by a factor $\xi/\Omega$, we
can safely drop them. This leads to
\begin{align}
 p_{\lambda\mu}^x =B_a\dot{x}_{\lambda\mu}, \quad
 p_{\lambda\mu}^y =B_b\dot{y}_{\lambda\mu}, \quad
 p_{\lambda\mu}^z =B_c\dot{z}_{\lambda\mu},
\end{align}
such that the canonical momenta of vibrations are directly
proportional to the velocities.

\subsection{Effective Hamiltonian}

Applying the usual Legendre transformations, the effective Hamiltonian is obtained as
\begin{align}\label{chp5:eq13}
 H&=\dot{\alpha}p_\alpha+\dot{\beta}p_\beta+\dot{\gamma}p_\gamma
   +\frac{1}{4\pi}\int d\Omega (\dot{x}p^x+\dot{y}p^y+\dot{z}p^z)-L\notag\\
  &=\frac{A_a}{2}\omega_x^2+\frac{A_b}{2}\omega_y^2+\frac{A_c}{2}\omega_z^2
   +\frac{B_a}{2}\sum_{\lambda\mu}\dot{x}_{\lambda\mu}^2
   +\frac{B_b}{2}\sum_{\lambda\mu}\dot{y}_{\lambda\mu}^2
   +\frac{B_c}{2}\sum_{\lambda\mu}\dot{z}_{\lambda\mu}^2\notag\\
  &\quad +\frac{1}{2}\sum_{\lambda\mu} \Big\{ \lambda(\lambda+1)\Big[D_a x_{\lambda\mu}^2
   +D_b y_{\lambda\mu}^2+D_c z_{\lambda\mu}^2\Big]+\mu^2\Big[D_d x_{\lambda\mu}^2
   +D_e y_{\lambda\mu}^2+D_f z_{\lambda\mu}^2\Big]\Big\},
\end{align}
with the rotational frequencies given by
\begin{align}
 \omega_x&=\frac{1}{A_a}\Big[I_1-\frac{B_a-2B_b}{2B_b}\sum_{\lambda\mu}z_{\lambda\mu}p_{\lambda\mu}^y
   -\frac{2B_c-B_a}{2B_c}\sum_{\lambda\mu}y_{\lambda\mu}p_{\lambda\mu}^z\Big],\\
 \omega_y&=\frac{1}{A_b}\Big[I_2-\frac{B_b-2B_c}{2B_c}\sum_{\lambda\mu}x_{\lambda\mu}p_{\lambda\mu}^z
   -\frac{2B_a-B_b}{2B_a}\sum_{\lambda\mu}z_{\lambda\mu}p_{\lambda\mu}^x\Big],\\
 \omega_z&=\frac{1}{A_c}\Big[I_3-\frac{B_c-2B_a}{2B_a}\sum_{\lambda\mu}y_{\lambda\mu}p_{\lambda\mu}^x
   -\frac{2B_b-B_c}{2B_b}\sum_{\lambda\mu}x_{\lambda\mu}p_{\lambda\mu}^y\Big],
\end{align}
and the vibrational velocities are directly proportional to the momenta
\begin{align}
 \dot{x}_{\lambda\mu}
 &=\frac{p_{\lambda\mu}^x}{B_a},\quad
 \dot{y}_{\lambda\mu}
  =\frac{p_{\lambda\mu}^y}{B_b},\quad
 \dot{z}_{\lambda\mu}
  =\frac{p_{\lambda\mu}^z}{B_c}~.
\end{align}
Substituting these relations into the Hamiltonian $H$,
one extracts the leading order part $\sim \Omega$
\begin{align}
 H_{\Omega}
  &=\sum_{\lambda\mu}\Big\{\frac{(p_{\lambda\mu}^x)^2}{2B_a}
   +\frac{(p_{\lambda\mu}^y)^2}{2B_b}+\frac{(p_{\lambda\mu}^z)^2}{2B_c}\notag\\
  &\quad +\frac{1}{2}\lambda(\lambda+1)\Big[D_a x_{\lambda\mu}^2
   +D_b y_{\lambda\mu}^2+D_c z_{\lambda\mu}^2\Big]
   +\frac{1}{2}\mu^2\Big[D_d x_{\lambda\mu}^2
   +D_e y_{\lambda\mu}^2+D_f z_{\lambda\mu}^2\Big] \Big\}~.
\end{align}
This Hamiltonian describes a set of infinitely many uncoupled (anisotropic)
harmonic oscillators with (vibrational) energies
\begin{align}
 \Omega_{\lambda\mu}^x &=\sqrt{[\lambda(\lambda+1)D_a+\mu^2 D_{d}]/B_a},\\
 \Omega_{\lambda\mu}^y &=\sqrt{[\lambda(\lambda+1)D_b+\mu^2 D_{e}]/B_b},\\
 \Omega_{\lambda\mu}^z &=\sqrt{[\lambda(\lambda+1)D_c+\mu^2 D_{f}]/B_c},
\end{align}
depending on the excitation mode $\lambda\mu$, the mass parameters $B_{a,b,c}$,
and the parameters of the potential $D_{a-f}$. Correspondingly, the energy
eigenvalues of this Hamiltonian are
\begin{align}\label{chp5:eq28}
 E_\Omega=\sum_{\lambda\mu} \Big[\Big(n_{\lambda\mu}^x+\frac{1}{2}\Big)\Omega_{\lambda\mu}^x
  +\Big(n_{\lambda\mu}^y+\frac{1}{2}\Big)\Omega_{\lambda\mu}^y
  +\Big(n_{\lambda\mu}^z+\frac{1}{2}\Big)\Omega_{\lambda\mu}^z\Big],
\end{align}
with oscillator quantum numbers $n_{\lambda\mu}^{x,y,z}=0, 1, 2, \dots$
and eigenfunctions given by products of cartesian oscillator
wave functions
\begin{align}
 |\Phi\rangle=\prod_{\lambda\mu}|n_{\lambda\mu}^x\rangle |n_{\lambda\mu}^y\rangle
  |n_{\lambda\mu}^z\rangle~.
\end{align}
The parity of such a state is $(-1)^{\sum_{\lambda\mu}(n_{\lambda\mu}^x
+n_{\lambda\mu}^y+n_{\lambda\mu}^z)}$, and since the ground state
of an even-even nucleus has positive parity, one requires that
$\sum_{\lambda\mu}(n_{\lambda\mu}^x+n_{\lambda\mu}^y+n_{\lambda\mu}^z)$
is even.

The next-to-leading order correction in the Hamiltonian $H$ is of order
$\xi$. It takes the form of a rotational Hamiltonian
\begin{align}\label{chp5:eq26}
 H_\xi = \frac{(I_1-l_1)^2}{2A_a} + \frac{(I_2-l_2)^2}{2A_b}
  +\frac{(I_3-l_3)^2}{2A_c},
\end{align}
with the collective angular momenta shifted by vibrational contributions
\begin{align}
 l_1&=\frac{B_a-2B_b}{2B_b}\sum_{\lambda\mu}z_{\lambda\mu}p_{\lambda\mu}^y
    +\frac{2B_c-B_a}{2B_c}\sum_{\lambda\mu}y_{\lambda\mu}p_{\lambda\mu}^z,\\
 l_2&=\frac{B_b-2B_c}{2B_c}\sum_{\lambda\mu}x_{\lambda\mu}p_{\lambda\mu}^z
    +\frac{2B_a-B_b}{2B_a}\sum_{\lambda\mu}z_{\lambda\mu}p_{\lambda\mu}^x,\\
 l_3&=\frac{B_c-2B_a}{2B_a}\sum_{\lambda\mu}y_{\lambda\mu}p_{\lambda\mu}^x
    +\frac{2B_b-B_c}{2B_b}\sum_{\lambda\mu}x_{\lambda\mu}p_{\lambda\mu}^y.
\end{align}
This implies that each vibrational state becomes a bandhead in
the rotational spectrum.

As a side remark, we note that nuclei with axial symmetry are realized by
the following parameters: $A_b=A_a$, $A_c=0$, $B_b=B_a$, $B_c=0$,
$D_b=D_a$, $D_c=0$, $D_e=D_d$, and $D_f=0$. In this case, the Hamiltonian
$H=H_\Omega+H_\xi$ simplifies drastically to
\begin{align}
 H_\Omega &=\sum_{\lambda\mu}\Big\{\frac{(p_{\lambda\mu}^x)^2+(p_{\lambda\mu}^y)^2}{2B_a}
   +\frac{1}{2}\Big[\lambda(\lambda+1)D_a+\mu^2 D_d\Big]\Big[x_{\lambda\mu}^2
   +y_{\lambda\mu}^2\Big]\Big\},\\
 H_\xi &=\frac{I_1^2+I_2^2}{2A_a}
        =\frac{I^2-I_3^2}{2A_a},
\end{align}
with
\begin{align}
 I_3=B_a\sum_{\lambda\mu}\Big(x_{\lambda\mu}\dot{y}_{\lambda\mu}
  -y_{\lambda\mu}\dot{x}_{\lambda\mu}\Big)
  =\sum_{\lambda\mu}\Big(x_{\lambda\mu}p_{\lambda\mu}^y-y_{\lambda\mu}p_{\lambda\mu}^x\Big),
\end{align}
consistent with the constructions in Refs.~\cite{Papenbrock2014PRC,
Papenbrock2015JPG, Papenbrock2016PS}.

Moreover, if the vibrational degrees of freedom are neglected, the
rotational Hamiltonian reads
\begin{align}
 H_\xi=\frac{I_1^2}{2A_a}+\frac{I_2^2}{2A_b}+\frac{I_3^2}{2A_c},
\end{align}
as derived for the triaxial rotor at leading order in
Ref.~\cite{Q.B.Chen2017EPJA}.

\section{Perturbative solution}\label{sec3}

As mentioned above, the vibrations generates contributions for the angular
momenta $I_1$, $I_2$, and $I_3$. With these contributions, the rotational
Hamiltonian $H_\xi$ in Eq.~(\ref{chp5:eq26}) becomes too complicated to be
solved exactly. In the following, we will use first order perturbation theory
to account for the vibrational corrections in $H_\xi$.

Let us start with the first term in $H_\xi$ proportional to
\begin{align}
 (I_1-l_1)^2=I_1^2 -2I_1l_1+l_1^2,
\end{align}
and take its expectation value in the vibrational state $|\Phi\rangle$.
Clearly, $I_1^2$ is not affected, since $|\Phi\rangle$ is
independent of the Euler angles. The second term $I_1l_1$ gives zero,
because the expectation value of a single position or momentum operator
vanishes
\begin{align}\label{chp5:eq27}
\langle n|x|n\rangle=0, \quad \langle n|p^x|n\rangle=0~.
\end{align}
Thus, we need to calculate only the expectation value of the third term
\begin{align}
 l_1^2
 &=\Big(\frac{B_a-2B_b}{2B_b}\Big)^2\sum_{\lambda\mu}\sum_{\lambda^\prime\mu^\prime}
  z_{\lambda\mu}p_{\lambda\mu}^y z_{\lambda^\prime\mu^\prime}p_{\lambda^\prime\mu^\prime}^y
  +\Big(\frac{2B_c-B_a}{2B_c}\Big)^2\sum_{\lambda\mu}\sum_{\lambda^\prime\mu^\prime}
  y_{\lambda\mu}p_{\lambda\mu}^z y_{\lambda^\prime\mu^\prime}p_{\lambda^\prime\mu^\prime}^z\notag\\
 &\quad +\Big(\frac{B_a-2B_b}{2B_b}\Big)\Big(\frac{2B_c-B_a}{2B_c}\Big)
  \sum_{\lambda\mu}\sum_{\lambda^\prime\mu^\prime}\Big(
  z_{\lambda\mu} p_{\lambda^\prime\mu^\prime}^z p_{\lambda\mu}^y y_{\lambda^\prime\mu^\prime}
  +y_{\lambda\mu}p_{\lambda^\prime\mu^\prime}^y p_{\lambda\mu}^z z_{\lambda^\prime\mu^\prime} \Big).
\end{align}
In the above double sums, the non-diagonal terms give still zero
according to Eq.~(\ref{chp5:eq27}). For the diagonal terms
\begin{align}
 &\quad \Big(\frac{B_a-2B_b}{2B_b}\Big)^2\sum_{\lambda\mu}
  z_{\lambda\mu}^2p_{\lambda\mu}^{y2}
  +\Big(\frac{2B_c-B_a}{2B_c}\Big)^2\sum_{\lambda\mu}
  y_{\lambda\mu}^2p_{\lambda\mu}^{z2}\notag\\
 &\quad +\Big(\frac{B_a-2B_b}{2B_b}\Big)\Big(\frac{2B_c-B_a}{2B_c}\Big)
  \sum_{\lambda\mu} \Big(z_{\lambda\mu}p_{\lambda\mu}^z~p_{\lambda\mu}^y y_{\lambda\mu}
  +y_{\lambda\mu}p_{\lambda\mu}^y~p_{\lambda\mu}^z z_{\lambda\mu}\Big),
\end{align}
the expectation values of squared positions, squared momenta, and products
of position and momentum of the same kind in a harmonic oscillator state
are easily computed. Altogether, the expectation value of $l_1^2$ in the state
$|\Phi\rangle$ is given by
\begin{align}
 \langle l_1^2 \rangle & =\frac{(B_a-2B_b)^2}{16B_bB_c}\sum_{\lambda\mu}
 \frac{\Omega_{\lambda\mu}^y}{\Omega_{\lambda\mu}^z}
  (2n_{\lambda\mu}^y+1)(2n_{\lambda\mu}^z+1)\hbar^2\notag\\
 &\quad +\frac{(2B_c-B_a)^2}{16B_bB_c}\sum_{\lambda\mu}
 \frac{\Omega_{\lambda\mu}^z}{\Omega_{\lambda\mu}^y}
  (2n_{\lambda\mu}^z+1)(2n_{\lambda\mu}^y+1)\hbar^2\notag\\
 &\quad +\frac{(B_a-2B_b)(2B_c-B_a)}{8B_bB_c}\sum_{\lambda\mu} \hbar^2.
\end{align}
The second and third term in $H_\xi$ are treated in the same way.
The inclusion of vibrational corrections to first order in the
rotational Hamiltonian $H_\xi$ leads to the result
\begin{align}
 \bar{H}_\xi &=\frac{1}{2A_a}\Big[I_1^2+\sum_{\lambda\mu} (l_{\lambda\mu}^1)^2\Big]
  +\frac{1}{2A_b}\Big[I_2^2+\sum_{\lambda\mu} (l_{\lambda\mu}^2)^2\Big]
  +\frac{1}{2A_c}\Big[I_3^2+\sum_{\lambda\mu} (l_{\lambda\mu}^3)^2\Big],
\end{align}
with the mode-specific angular momentum shifts
\begin{align}
 l_{\lambda\mu}^1 &=\Big[\frac{(B_a-2B_b)^2}{16B_bB_c}
 \frac{\Omega_{\lambda\mu}^y}{\Omega_{\lambda\mu}^z}
  (2n_{\lambda\mu}^y+1)(2n_{\lambda\mu}^z+1)
  +\frac{(2B_c-B_a)^2}{16B_bB_c}
 \frac{\Omega_{\lambda\mu}^z}{\Omega_{\lambda\mu}^y}
  (2n_{\lambda\mu}^z+1)(2n_{\lambda\mu}^y+1)\notag\\
  &\quad +\frac{(B_a-2B_b)(2B_c-B_a)}{8B_bB_c} \Big]^{1/2}\hbar, \notag\\
 l_{\lambda\mu}^2 &=\Big[\frac{(B_b-2B_c)^2}{16B_cB_a}
 \frac{\Omega_{\lambda\mu}^z}{\Omega_{\lambda\mu}^x}
  (2n_{\lambda\mu}^z+1)(2n_{\lambda\mu}^x+1)
  +\frac{(2B_a-B_b)^2}{16B_cB_a}
 \frac{\Omega_{\lambda\mu}^x}{\Omega_{\lambda\mu}^z}
  (2n_{\lambda\mu}^x+1)(2n_{\lambda\mu}^z+1)\notag\\
  &\quad +\frac{(B_b-2B_c)(2B_a-B_b)}{8B_cB_a} \Big]^{1/2}\hbar, \notag\\
 l_{\lambda\mu}^3 &=\Big[\frac{(B_c-2B_a)^2}{16B_aB_b}
 \frac{\Omega_{\lambda\mu}^x}{\Omega_{\lambda\mu}^y}
  (2n_{\lambda\mu}^x+1)(2n_{\lambda\mu}^y+1)
  +\frac{(2B_b-B_c)^2}{16B_aB_b}
 \frac{\Omega_{\lambda\mu}^y}{\Omega_{\lambda\mu}^x}
  (2n_{\lambda\mu}^y+1)(2n_{\lambda\mu}^x+1)\notag\\
 &\quad +\frac{(B_c-2B_a)(2B_b-B_c)}{8B_aB_b}\Big]^{1/2}\hbar.
\end{align}
This shows that the angular momentum contributions from the vibrational
motion play the role of recoil terms~\cite{Bohr1975, Ring1980book} in
first order perturbation theory. For different vibrational states,
the spin components of the bandhead are different and characterized
by the vibrational quantum numbers $n_{\lambda\mu}^x$, $n_{\lambda\mu}^y$, and
$n_{\lambda\mu}^z$.

We can now specify the corrections for each vibrational state.
For the ground state with quantum numbers
$n_{\lambda\mu}^x=n_{\lambda\mu}^y=n_{\lambda\mu}^z=0$ and
energy eigenvalue $E_\Omega=\frac{1}{2}\sum_{\lambda\mu} \big[\Omega_{\lambda\mu}^x
  +\Omega_{\lambda\mu}^y +\Omega_{\lambda\mu}^z\big]$, one has
the angular momentum shifts
\begin{align}\label{chp5:eq30}
 l_{\lambda\mu}^1 &=\frac{\hbar}{4\sqrt{B_bB_c}}\Big|
  (B_a-2B_b)\sqrt{\frac{\Omega_{\lambda\mu}^y}{\Omega_{\lambda\mu}^z}}
  +(2B_c-B_a)\sqrt{\frac{\Omega_{\lambda\mu}^z}{\Omega_{\lambda\mu}^y}}\Big|,\notag\\
 l_{\lambda\mu}^2 &=\frac{\hbar}{4\sqrt{B_cB_a}}\Big|
  (B_b-2B_c)\sqrt{\frac{\Omega_{\lambda\mu}^z}{\Omega_{\lambda\mu}^x}}
  +(2B_a-B_b)\sqrt{\frac{\Omega_{\lambda\mu}^x}{\Omega_{\lambda\mu}^z}}\Big|,\notag\\
 l_{\lambda\mu}^3 &=\frac{\hbar}{4\sqrt{B_aB_b}}\Big|
  (B_c-2B_a)\sqrt{\frac{\Omega_{\lambda\mu}^x}{\Omega_{\lambda\mu}^y}}
  +(2B_b-B_c)\sqrt{\frac{\Omega_{\lambda\mu}^y}{\Omega_{\lambda\mu}^x}}\Big|.
\end{align}

Next, we consider the excitation of one particular mode $\lambda\mu$.
Assuming the ordering of energies $\Omega_{\lambda\mu}^x<\Omega_{\lambda\mu}^y<\Omega_{\lambda\mu}^z$,
the lowest vibrational excitation with positive parity has quantum
numbers $n_{\lambda\mu}^x=2$, $n_{\lambda\mu}^y=n_{\lambda\mu}^z=0$.
In this case, the excitation energy is $2\Omega_{\lambda\mu}^x$
and the angular momentum shifts are
\begin{align}
 l_{\lambda\mu}^1 &= \Big[\frac{(B_a-2B_b)^2}{16B_bB_c}
 \frac{\Omega_{\lambda\mu}^y}{\Omega_{\lambda\mu}^z}
  +\frac{(2B_c-B_a)^2}{16B_bB_c}
 \frac{\Omega_{\lambda\mu}^z}{\Omega_{\lambda\mu}^y}
  +\frac{(B_a-2B_b)(2B_c-B_a)}{8B_bB_c} \Big]^{1/2}\hbar,\notag\\
 l_{\lambda\mu}^2 &= \Big[\frac{(B_b-2B_c)^2}{16B_cB_a}
 \frac{5\Omega_{\lambda\mu}^z}{\Omega_{\lambda\mu}^x}
  +\frac{(2B_a-B_b)^2}{16B_cB_a}
 \frac{5\Omega_{\lambda\mu}^x}{\Omega_{\lambda\mu}^z}
  +\frac{(B_b-2B_c)(2B_a-B_b)}{8B_cB_a} \Big]^{1/2}\hbar,\notag\\
 l_{\lambda\mu}^3 &= \Big[\frac{(B_c-2B_a)^2}{16B_aB_b}
 \frac{5\Omega_{\lambda\mu}^x}{\Omega_{\lambda\mu}^y}
  +\frac{(2B_b-B_c)^2}{16B_aB_b}
 \frac{5\Omega_{\lambda\mu}^y}{\Omega_{\lambda\mu}^x}
  +\frac{(B_c-2B_a)(2B_b-B_c)}{8B_aB_b} \Big]^{1/2}\hbar.
\end{align}
One can see that in comparison with that for the ground state
band Eq.~(\ref{chp5:eq30}), the recoil terms are different in the
excited states.

\section{Expressions in terms of quadrupole deformations}\label{sec4}

In the construction of the effective Lagrangian (\ref{chp5:eq43}),
the vibrational degrees of freedom were denoted by $x$, $y$, and $z$.
Its rotational part (proportional to $A_{a,b,c}$) involves vibrational
contributions to the rotational frequencies and thus became too complicated.
Following  empirical experience, we express $x$, $y$, and $z$
in terms of the conventional quadrupole deformation parameters
$\beta_2$ and $\gamma_2$ of a triaxially deformed nucleus.

Starting from the equation of the nuclear surface~\cite{Ring1980book}
\begin{align}
 R(\theta,\phi)=R_0\Big\{1+\beta_2\Big[\cos\gamma_2 Y_{20}(\theta,\phi)
 +\frac{\sin\gamma_2}{\sqrt{2}}\Big(Y_{22}(\theta,\phi)+Y_{2-2}(\theta,\phi)\Big)\Big] \Big\},
\end{align}
with $R_0=1.2A^{1/3}~\textrm{fm}$, the displacements $x$, $y$, and $z$
take the form
\begin{align}
 x&=[R(\theta,\phi)-R_0]\sin\theta\cos\phi,\\
 y&=[R(\theta,\phi)-R_0]\sin\theta\sin\phi,\\
 z&=[R(\theta,\phi)-R_0]\cos\theta~.
\end{align}
By making use of the relations
\begin{align}
 \cos\theta~Y_{\lambda\mu} &=a_{\lambda,\mu}Y_{\lambda+1,\mu}+a_{\lambda-1,\mu}Y_{\lambda-1,\mu},\\
 \sin\theta e^{i\phi}~Y_{\lambda\mu} &=b_{\lambda-1,-(\mu+1)}Y_{\lambda-1,\mu+1}-b_{\lambda,\mu}Y_{\lambda+1,\mu+1},\\
 \sin\theta e^{-i\phi}~Y_{\lambda\mu} &=-b_{\lambda-1,\mu-1}Y_{\lambda-1,\mu-1}+b_{\lambda,-\mu}Y_{\lambda+1,\mu-1},
\end{align}
with the coefficients
\begin{align}
 a_{\lambda\mu}=\sqrt{\frac{(\lambda+1)^2-\mu^2}{(2\lambda+1)(2\lambda+3)}},\quad
 b_{\lambda\mu}=\sqrt{\frac{(\lambda+\mu+1)(\lambda+\mu+2)}{(2\lambda+1)(2\lambda+3)}},
\end{align}
the few non-vanishing expansion coefficients $x_{\lambda\mu}$, $y_{\lambda\mu}$,
and $z_{\lambda\mu}$ defined by Eq.~(\ref{chp5:eq31}) are
given by
\begin{align}
\label{chp5:eq32}
 x_{31}&=-\sqrt{\frac{6}{35}}R_0\beta_2\cos\gamma_2
 +\sqrt{\frac{1}{70}}R_0\beta_2\sin\gamma_2, \quad
 x_{33}=-\sqrt{\frac{3}{14}}R_0\beta_2\sin\gamma_2,\\
\label{chp5:eq33}
 y_{3-1}&=-\sqrt{\frac{6}{35}}R_0\beta_2\cos\gamma_2
 -\sqrt{\frac{1}{70}}R_0\beta_2\sin\gamma_2,\quad
 y_{3-3}=-\sqrt{\frac{3}{14}}R_0\beta_2\sin\gamma_2,\\
\label{chp5:eq34}
 z_{30}&=\sqrt{\frac{9}{35}}R_0\beta_2\cos\gamma_2,\quad
 z_{32}=\sqrt{\frac{1}{7}}R_0\beta_2\sin\gamma_2.
\end{align}
With these restricted modes, the vibrational contributions to the angular momenta
in Eqs.~(\ref{chp5:eq44})-(\ref{chp5:eq45}) actually vanish.

Substituting the modes (\ref{chp5:eq32})-(\ref{chp5:eq34}) into the
Lagrangian (\ref{chp5:eq19}), calculating the corresponding canonical
momenta, and performing the Legendre transformation, one
arrives at the following LO Hamiltonian (vibrational part)
\begin{align}
 H_\Omega
 &=\frac{1}{2}B_{\beta \beta }\,\dot{\beta}_2^2
  +\frac{1}{2}B_{\gamma\gamma}\,\beta_2^2\dot{\gamma}_2^2
  +B_{\beta\gamma}\,\beta_2 \dot{\beta}_2\dot{\gamma}_2+V,
\end{align}
with the collective potential
\begin{align}\label{chp5:eq36}
 V(\beta_2,\gamma_2)&=\frac{3}{35} R_0^2\big(12D_a+12D_b+18D_c+D_d+D_e\big)\beta_2^2\cos^2\gamma_2\notag\\
 &\quad +\frac{2}{35}R_0^2\big(24D_a+24D_b+15D_c+17D_d+17D_e+5D_f\big)\beta_2^2\sin^2\gamma_2\notag\\
 &\quad +\frac{\sqrt{3}}{35} R_0^2
 \big(12D_b+D_e-12D_a-D_d\big)\beta_2^2 \sin\gamma_2\cos\gamma_2~.
\end{align}
Furthermore, the mass parameters read
\begin{align}\label{chp5:eq37}
 B_{\beta\beta} &=\frac{R_0^2}{35}\Big[B_a\Big(7-2\sin(2\gamma_2+\frac{\pi}{6})\Big)
  +B_b\Big(7+2\sin(2\gamma_2-\frac{\pi}{6})\Big)
  +7B_c\Big(1+1\cos 2\gamma_2\Big)\Big],\notag\\
 B_{\gamma\gamma}&=\frac{R_0^2}{35}\Big[B_a\Big(7+2\sin(2\gamma_2+\frac{\pi}{6})\Big)
  +B_b\Big(7-2\sin(2\gamma_2-\frac{\pi}{6})\Big)
  +7B_c\Big(1-1\cos2\gamma_2\Big)\Big],\notag\\
 B_{\beta\gamma}&=\frac{R_0^2}{35}\Big[2B_a\sin(2\gamma_2-\frac{\pi}{3})
  +2B_b\sin(2\gamma_2+\frac{\pi}{3})
  -2B_c\sin2\gamma_2 \Big],
\end{align}
and interestingly, they do not depend on $\beta_2$.

Moreover, the rotational part of the NLO Hamiltonian  is just
\begin{align}
 H_{\xi}=\frac{I_1^2}{2A_a}+\frac{I_2^2}{2A_b}+\frac{I_3^2}{2A_c},
\end{align}
with moments of inertia $A_a$, $A_b$, and $A_c$ independent of the
deformation parameters $\beta_2$ and $\gamma_2$.

The total Hamiltonian
\begin{align}\label{chp5:eq35}
 H=H_\Omega+H_\xi,
\end{align}
is a rotation-vibration Hamiltonian of the Bohr-Mottelson type~\cite{Bohr1975}
with definite potential $V$ and non-constant mass parameters. In the following, this
Hamiltonian will be used to describe the experimental energy
spectra.

\section{Results and Discussions}\label{sec5}

In this section, the experimental ground state bands, $\gamma$-bands, and $K=4$
bands for the isotopes $^{108,110,112}$Ru are used to examine
the applicability of the present EFT in the description of collective
rotations and vibrations of triaxially deformed nuclei. The data are
taken from the compilation of the National Nuclear Data Center
(NNDC)~\footnote{http://www.nndc.bnl.gov/ensdf/.}. In
Ref.~\cite{Q.B.Chen2017EPJA}, it has been shown that these three Ru
isotopes have a triaxially deformed minimum and exhibit
softness along the $\gamma_2$-direction in the potential energy
surface, calculated by CDFT~\cite{J.Meng2016book}.
Moreover, the nearly constant behavior of the experimental
alignments in the spin region of $I\leq 10\hbar$
indicates that the corresponding data are not beyond
the breakdown energy scale for collective rotational and
vibrational motions (i.e., beyond the pairing
instability)~\cite{Q.B.Chen2017EPJA}. Therefore, the application
of the rotation-vibration Hamiltonian (\ref{chp5:eq35}) is restricted
to this spin region.

\begin{figure*}[h]
  \begin{center}
    \includegraphics[width=15.0 cm]{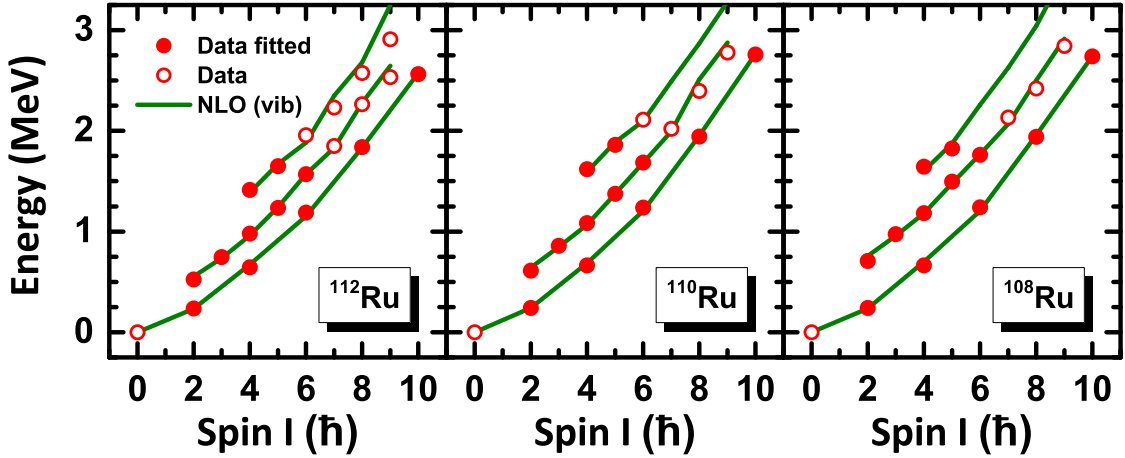}
    \caption{Energy spectra for the ground state, $\gamma$-, and $K=4$ bands
    in $^{108-112}$Ru calculated by rotation-vibration Hamiltonian (\ref{chp5:eq35}).}\label{fig1}
  \end{center}
\end{figure*}

In Fig.~\ref{fig1}, the energy spectra of the ground state bands,
$\gamma$-bands, and $K = 4$ bands obtained from rotation-vibration
Hamiltonian (\ref{chp5:eq35}) are shown as a function of
spin $I$ for the isotopes $^{108, 110, 112}$Ru, respectively. The parameters
of the Hamiltonian are obtained by fitting to the solid data points
in Fig.~\ref{fig1}, and their values are listed in
Tab.~\ref{tab1}. One can see that the data are well reproduced by
the rotation-vibration Hamiltonian (\ref{chp5:eq35}). In particular,
the obvious staggering behavior of the $\gamma$-bands found with the
triaxial rotor Hamiltonian~\cite{Q.B.Chen2017EPJA} is no longer present.
This improved description underlines the importance of including
vibrational degrees of freedom in the EFT formulation.

From the moments of inertia, mass parameters, and potential parameters
collected in Tab.~\ref{tab1}, one recognizes significant differences
between the isotope $^{108}$Ru and $^{112,110}$Ru. This is
consistent with the fact that the energy differences between the ground state
and $\gamma$- bands in $^{108}$Ru ($\sim 500~\textrm{keV}$) are larger
than those in $^{112,110}$Ru ($\sim 350~\textrm{keV}$). This also indicates that
one has to fit the parameters for each nucleus separately, and this, to some extent,
weakens the predictive power of the EFT.

\begin{table*}[!ht]
\caption{Parameters used in the rotation-vibration Hamiltonian (\ref{chp5:eq35})
 for calculations of $^{108-112}$Ru. The units of $A_{a,b,c}$ are $\hbar^2/\textrm{MeV}$,
 the units of $B_{a,b,c}$ are $\hbar^2/(\textrm{MeV}\cdot {\textrm{fm}}^2)$, and
 the units of $D_{a-f}$ are $\textrm{MeV}/\textrm{fm}^2$.} \label{tab1}
\begin{tabular}{rrrrrrrrrrrrrrrrrr}
\hline
\hline
Nucleus & $A_a~$ & $A_b~$   & $A_c~$ & $B_a~$ & $B_b~$ & $B_c~$
 & $D_a~$ & $D_b~$ & $D_c~$ & $D_d~$ & $D_e~$ & $D_f~$ \\
\hline
$^{112}$Ru & 15.0  &   26.7  &  17.8 &  $24.3$ &  99.5  &  0.02  &  34.3
 &  $-20.4$  &  $-6.1$  &  65.8  &  $-42.2$  &  81.9   \\
$^{110}$Ru & 15.7  & 24.0 & 12.0 &  $17.8$ &  29.0  &  0.04  &  26.6
 &  $-14.5$  &  $-5.5$  &   24.0  &  $-11.1$  &  57.7 \\
$^{108}$Ru & $15.1$  &  $24.9$  &  $18.0$  &  $~6.5$ &  $3.0$ &  $0.01$
 &  $99.0$  & $-65.0$  &  $-11.3$  & 264.5 & $-207.3$ &  259.9\\
\hline
\end{tabular}
\end{table*}

\begin{figure*}[h]
  \begin{center}
    \vspace{-1.5 cm}
    \includegraphics[width=15.0 cm]{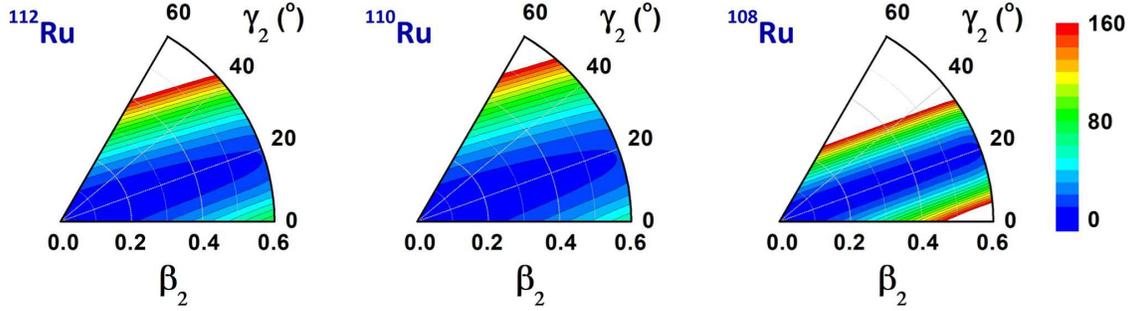}
    \vspace{-2.0 cm}
    \caption{Collective potentials for $^{108-112}$Ru calculated by Eq.~(\ref{chp5:eq36}).}\label{fig2}
  \end{center}
\end{figure*}

\begin{figure*}[h]
  \begin{center}
    \includegraphics[width=15.0 cm]{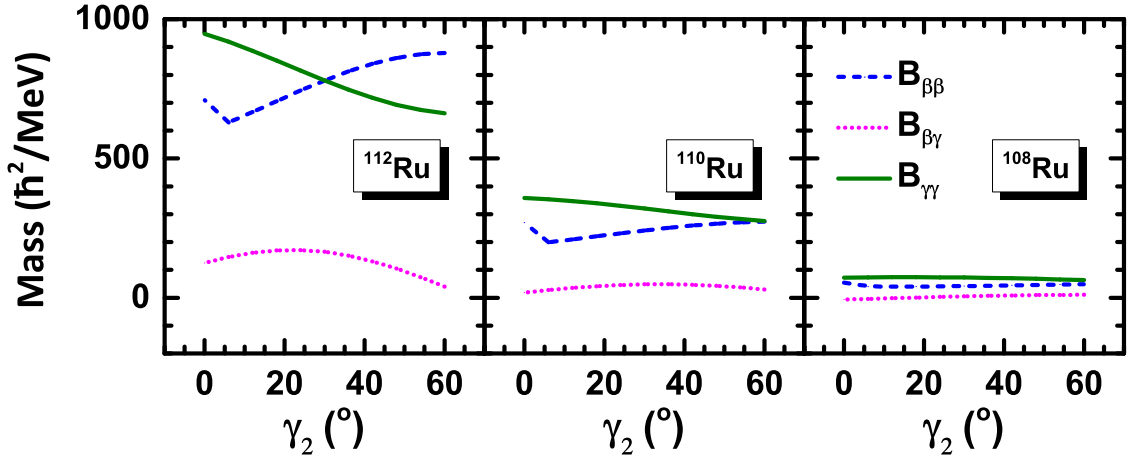}
    \caption{Mass parameters for $^{108-112}$Ru calculated by Eq.~(\ref{chp5:eq37}).}\label{fig3}
  \end{center}
\end{figure*}

With the parameters listed in Tab.~\ref{tab1}, the collective
potential $V(\beta_2,\gamma_2)$ (\ref{chp5:eq36}) and the mass parameters
$B_{\beta\beta}$, $B_{\beta\gamma}$, and $B_{\gamma\gamma}$ (\ref{chp5:eq37}) are
determined and these are shown for the three Ru isotopes in Figs.~\ref{fig2}
and~\ref{fig3}, respectively. One observes that
the potential energy surfaces shown in Fig.~\ref{fig2}
possess similar shapes. Namely, there is a spherical minimum and
softness along the $\beta_2$-direction with a valley located
around $\gamma_2=20^\circ$. It should be noted that such
behavior of the potential energy surface is different from
those calculated by CDFT~\cite{Q.B.Chen2017EPJA, J.Meng2016book}
using the effective interaction PC-PK1~\cite{P.W.Zhao2010PRC},
in which a triaxially deformed minimum and softness along
the $\gamma_2$-direction are found. The differences may be due to
the fact that only terms proportional to $\beta_2^2$
can arise in the present LO collective
potential (\ref{chp5:eq36}). In order to get a more structured
shape of the potential energy surface of CDFT, the higher order terms
$\mathcal{L}_{5i}$ and $\mathcal{L}_{6i}$ in Eq.~(\ref{chp5:eq43})
need to be kept. This also indicates that the EFT formulation
gives a different picture for the descriptions of the energy spectra
of Ru isotopes in comparison to the five-dimensional collective
Hamiltonian (5DCH) based on CDFT~\cite{Q.B.Chen2017EPJA}.

We have already mentioned that the mass parameters $B_{\beta\beta}$, $B_{\beta\gamma}$,
and $B_{\gamma\gamma}$ are independent of the deformation $\beta_2$.
Furthermore, their dependence on $\gamma_2$ is moderate, as can be seen
in Fig.~\ref{fig3}. With increasing $\gamma_2$, $B_{\beta\beta}$ increases
and $B_{\gamma\gamma}$ decreases. One can observe that $B_{\beta\gamma}$
is much smaller than $B_{\beta\beta}$ and $B_{\gamma\gamma}$
for all three Ru isotopes, which implies that the coupling between
$\beta_2$ and $\gamma_2$ is small. Once again, we should point
out that the EFT formulation gives a different picture for collective
energy spectra in comparisons to those of the 5DCH based on
CDFT~\cite{Q.B.Chen2017EPJA}.

\section{Summary}\label{sec6}

In summary, the EFT for triaxially deformed even-even nuclei has been
generalized to include the vibrational degrees of freedom. The pertinent
Lagrangian and Hamiltonian were obtained up to NLO. The LO Hamiltonian
describes a set of uncoupled (anisotropic) harmonic oscillators.
The NLO part couples rotations to vibrations, and it is found that
the vibrations provide contributions to the angular momenta $I_1$, $I_2$, and
$I_3$. This coupling makes the rotational Hamiltonian too complicated to be solved
exactly.

Therefore, we have treated the NLO (rotational) Hamiltonian in first
order perturbation theory. This leads to corrections from the vibrational
motion in the form of so-called recoil term. For different vibrational
states, the spin components of the bandhead become different and they
depend on different vibrational quantum numbers.

The NLO Hamiltonian has also been expressed in terms of quadrupole
deformation parameters $\beta_2$ and $\gamma_2$. A rotation-vibration
Hamiltonian (without mutual coupling) is obtained. Its applicability
has been examined in the description of the energy spectra of
the ground state bands, $\gamma$-bands, and $K=4$ bands in $^{108,110,112}$Ru
isotopes. It is found that by taking into account the vibrational degree of freedom,
the deviations for high-spin states in the $\gamma$-band, using the EFT with
only rotational degree of freedom, disappear. This underlines the
importance of including vibrational degrees of freedom in the EFT formulation.

The results presented in this work give us confidence to further
generalize the EFT for triaxially deformed nuclei with odd mass
number, which requires a systematic treatment of the coupling
between the single particle motion and the collective rotational
motion.

\section*{Acknowledgements}

The authors thank T.~Papenbrock, P.~Ring, and W.~Weise for
helpful and informative discussions. This work was supported in part by the Deutsche
Forschungsgemeinschaft (DFG) and National Natural Science Foundation
of China (NSFC) through funds provided to the Sino-German CRC 110
``Symmetries and the Emergence of Structure in QCD'',
the Major State 973 Program of China No.~2013CB834400, and the
NSFC under Grants No.~11335002, and No.~11621131001. The work of
UGM was also supported by the Chinese Academy of Sciences (CAS)
President's International Fellowship Initiative (PIFI) (Grant No.~2018DM0034).

\end{document}